# Metamagnetism and 1/3 Plateau in the Spin Chain Compound $CoV_2O_6$

Simon A.J. Kimber[1,2,3], Dimitri N. Argyriou[1], J. Paul Attfield[2,3]

[1]*Hahn-Meitner Institute, 100 Glienicker Straße, 14109 Berlin, Germany*

[2]*Centre for Science at Extreme Conditions, University of Edinburgh, Erskine Williamson Building, King's Buildings, Mayfield Road, Edinburgh EH9 3JZ, United Kingdom*

[3]*School of Chemistry, University of Edinburgh, Joseph Black Building, King's Buildings, West Mains Road, Edinburgh EH9 3JJ, United Kingdom*

The pseudo-one dimensional brannerite type compound $CoV_2O_6$ has been studied by magnetisation and heat capacity measurements in the temperature range $1.8 < T < 300$ K and in applied fields of up to 9 T. Our measurements show an unusual balance of exchange interactions ($\Theta = 2.8(2)$ K) and a single antiferromagnetic transition at $T_N = 7$ K. M(H) isotherms recorded below 5 K show a metamagnetic transition and a 1/3 magnetisation plateau. From 5 – 7 K, only the metamagnetic transition is observed. For $T < T_N < \sim 25$ K, our heat capacity and magnetisation measurements show evidence for strong low dimensional ferromagnetic fluctuations. We propose a simple phase diagram, discuss the principle features and emphasise the importance of Ising anisotropy.

The emergence of magnetisation plateaus and other field dependant phenomena such as colossal magnetoresistances in transition metal oxides is of long standing interest [1]. Magnetic ions with strong Ising single ion anisotropy often show particularily rich behaviour, especially when combined with reduced structural dimensonality. A good example is the quasi-one dimensional Ising ferromagnet $CoNb_2O_6$ which has been the subject of numerous studies. The first neutron powder diffraction experiments found a transition to a incommensurately modulated magnetic structure at 2.9 K followed by a transition to a non-collinear antiferromagnetic phase at 1.9 K [2]. Magnetisation measurements made on powder samples at 1.4 K, reveal metamagnetic behaviour and a plateau at 1/3 of the saturation magnetisation [3]. The heat capacity of $CoNb_2O_6$ is highly anisotropic [4] and single crystal neutron diffraction shows strongly (*hkl*) dependent coherence lengths implying two dimensional order [5]. Numerous neutron diffraction studies in applied fields have reported a wealth of distinct field induced magnetic structures and the presence of a Lifshitz point (three phase co-existence) in the field/temperature field diagram [6-11]. The unusual low temperature properties of $CoNb_2O_6$ have been attributed to the combination of Ising anisotropy and magnetic frustration arising from the zig-zag arangement of $CoO_6$ octahedra.

In this paper we report our observations on the related material $CoV_2O_6$, which crystallises in the triclinic (P-1) brannerite structure shown in Fig. 1. The structure of $CoV_2O_6$ consists of well separated edge-sharing chains of $CoO_6$ octahedra running down the *b* axis interspersed by non-magnetic $V^{5+}$ in $V_2O_5$ like blocks [12]. In contrast to $CoNb_2O_6$, magnetic frustration is not expected to be important due to the linear connectivity of the $Co^{2+}$ spins. The magnetic properties of several other brannerites have been reported, and appear to be largely determined by single ion properties. The isostructural material $CuV_2O_6$ is a low dimensional antiferromagnet, due to $Cu^{2+}$ orbital order [13,14], whilst monoclinic $MnV_2O_6$ is an isotropic antiferromagnet that shows reduced magnetic coherence lengths due to antisite disorder [15]. Here we show that in common with other brannerites, $CoV_2O_6$ is antiferromagnetic, with a Néel transition at $T_N$ = 7 K. However, in contrast to the other brannerites, magnetisation measurements show metamagnetic behaviour and the presence of a 1/3 magnetisation plateau at low temperatures. Furthermore, our magnetisation and heat capacity measurements show evidence for ferromagnetic correlations for $T > T_N$. We have constructed a schematic phase diagram for $CoV_2O_6$ based on our measurements, and emphasise the importance of Ising anisotropy in determining the low temperature magnetic properties.

Polycrystalline $CoV_2O_6$ was synthesised using a citrate decomposition technique as described previously for $MnV_2O_6$ [15], powder X-ray diffraction showed a phase pure product. The magnetic susceptibilities for $CoV_2O_6$, measured in a 500 Oe field using a Quantum Designs SQUID magnetometer, are shown in Fig. 2a and show a sharp transition to an antiferromagnetically ordered state below $T_N$ = 7 K. No divergence between field and zero field cooled measurements was seen. The inverse susceptibility of $CoV_2O_6$ is well fitted by the Curie-Weiss law in the range 125 – 300 K with a fitted moment of 5.22(2) $\mu_B$ and a Weiss temperature of 2.8(2) K. The moment shows a substantial increase due to spin orbit coupling from the spin only value for S = 3/2 (3.87 $\mu_B$) as is often found for high spin Co(II) in octahedral coordination with a $^4T_{1g}$ ground state. The unusually small Weiss temperature could indicate weak magnetic exchange, or the near cancellation of strong ferro- and antiferromagnetic interactions. The plot of $\chi T$ vs. T shown in Fig. 2b, provides strong evidence for the latter view, as a sizeable enhancement in $\chi T$ is observed below ~40 K, before an overall antiferromagnetic state is realised at $T_N$. Hence strong ferromagnetic correlations, presumably originating in the $CoO_6$ chains are present well above $T_N$ and long range order is the result of antiferromagnetic interchain exchange. Additional susceptibility measurements were performed in increments of 500 Oe in order to map out the phase boundaries (see below)

M(H) isotherms were recorded at a range of temperatures in the magnetically ordered region and above $T_N$ in fields of up to 9 T using a Quantum Designs PPMS. Isotherms measured in the range $2 \leq T < T_N$ are shown in Fig. 3a. At 2 K, two field induced transitions are seen. The first transition at 3600 Oe corresponds to a plateau at almost exactly one third of the saturation magnetisation (0.95 $\mu_B$), the second transition at 5900 Oe is a metamagnetic transition to a plateau that shows the full saturation magnetisation, 2.9 $\mu_B$ at 9 T ($\mu_{sat} = 2S = 3\mu_B$). Considerable hysteresis was observed around the 1/3 plateau at the lowest temperatures measured. M(H) isotherms at higher temperatures show that the 1/3 plateau becomes non-hysteretic at 4 K and disappears between 4.5 and 5 K (Fig. 3a) whilst the metamagnetic transition persists up to the Néel temperature. Unusually, above $T_N$, M(H) isotherms characteristic of a soft ferromagnet are observed, (Fig. 3b), corroborating the evidence from the susceptibility measurements for ferromagnetic correlations in the $CoO_6$ edge sharing chains.

The emergence of the 1/3 magnetisation plateau at T < ~ 5 K could conceivably be the result of an additional low temperature magnetic transition, undetected in the susceptibility measurements. We therefore measured the heat capacity of a polycrystalline pellet of $CoV_2O_6$ from 2 – 75 K using a Quantum Designs PPMS (Fig. 3).

A sharp lambda anomaly is seen at 6.3 K, showing that only one magnetic transition occurs in this temperature range. In addition to being a sensitive probe of phase transitions, the magnetic entropy extracted from heat capacity measurements is also an excellent measure of the degree of 'low dimensionality' in magnetic systems. The lattice contribution to the heat capacity of $CoV_2O_6$ was estimated by fitting the high temperature (T > 30 K) heat capacity to a function of the form $BT^3 + CT^5$ and is shown as a solid line in the inset to Fig 3. The magnetic entropy was obtained by subtracting the lattice contribution and is shown as a bold line in Fig. 3. The total magnetic entropy we obtain at 50 K (5.43 J/mol.K) is very close to the expected value for an effective S = 1/2 spin (Rln2 = 5.76 J/mol.K). However, 65 % of the entropy is acquired above $T_N$, confirming the low dimensional nature of this system.

By combining the results of our measurements, we propose the schematic phase diagram shown in Fig. 5. Of particular interest is the 1/3 magnetisation plateau, which is previously unreported in the $MV_2O_6$ materials. Although fractional magnetisation plateaus are frequently discussed within the framework of frustrated spin systems [16], many unfrustrated Ising metamagnets also show such features. Given the unfrustrated topology of $Co^{2+}$ spins in $CoV_2O_6$, simple model metamagnets may be a good analogy. Well known examples include the $MCl_2.2H_2O$ (M = Fe, Co) materials. The phase diagrams of these compounds are very similar to that of $CoV_2O_6$, showing a triple point where the phase boundaries for the paramagnetic, 1/3 plateau and antiferromagnetic regions meet [17-21]. The $MCl_2.2H_2O$ materials also shown low Weiss constants (1±1 K for M = Co) [22]. However, these compounds do not show the unusual persistence of ferromagnetism above $T_N$ that we report for $CoV_2O_6$.

In conclusion, we have investigated the brannerite type material $CoV_2O_6$ by magnetic susceptibility and heat capacity measurements. Our results show that $CoV_2O_6$ is metamagnetic with a 1/3 magnetisation plateau in contrast to the magnetic properties of other reported brannerites. Neutron scattering investigations of $CoV_2O_6$ are in progress and will be reported in a future publication.

S.A.J.K and J.P.A thank the E.P.S.R.C and Leverhulme trust respectively for support.

FIG. 1 Structure of $CoV_2O_6$ projected down [001] showing edge sharing $CoO_6$ chains, hatched octahedra are $CoO_6$, plain polyhedra are $VO_5$.

FIG. 2 a) Magnetic susceptibility and inverse susceptibility of $CoV_2O_6$ measured in a field of 500 Oe and plotted as a function of temperature. Line shows Curie-Weiss fit to range 50 – 300 K extrapolated to low temperature; b) Plot of $\chi T$ vs. T, showing crossover from ferromagnetic fluctuations at high temperature to antiferromagnetic order at $T_N = 7$ K.

FIG. 3 a) M(H) isotherms measured on $CoV_2O_6$ for $T < T_N$ illustrating 1/3 plateau and metamagnetic transition, succesive curves have been displaced by 1500 Oe for clarity; b) M(H) isotherms measured on $CoV_2O_6$ for $T > T_N$, showing persistence of ferromagnetic fluctuations.

FIG. 4 Magnetic heat capacity of $CoV_2O_6$ measured in zero applied field, bold line shows integrated magnetic entropy. Inset shows total heat capacity, line shows calculated lattice contribution.

FIG. 5 Field/temperature phase diagram proposed for $CoV_2O_6$ from magnetic susceptibility measurements, hatched region shows extent of hysteresis at 1/3 plateau. Lines are guides to the eye.

1)

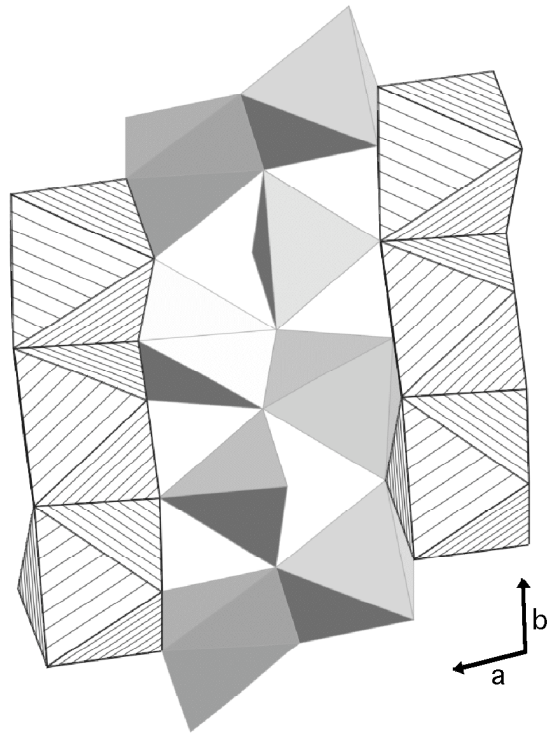

2)

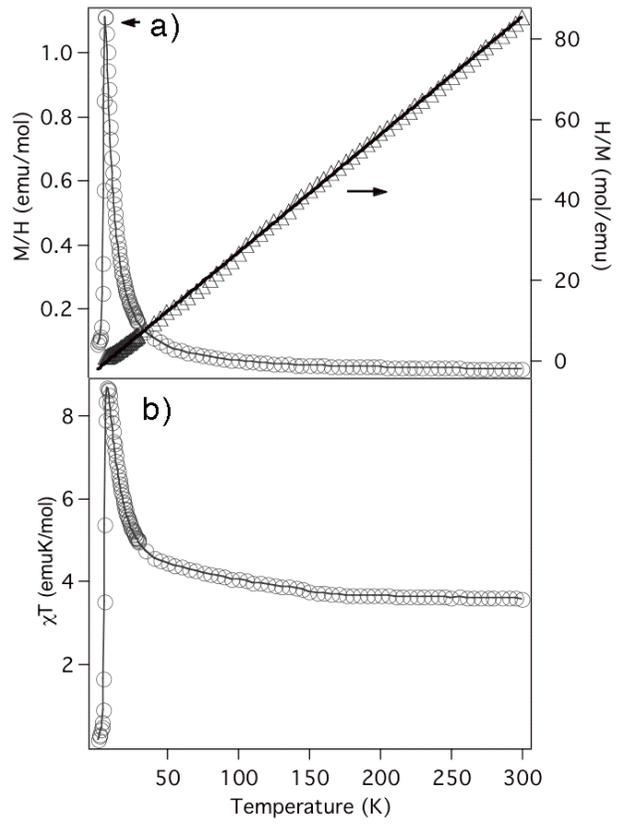

3)

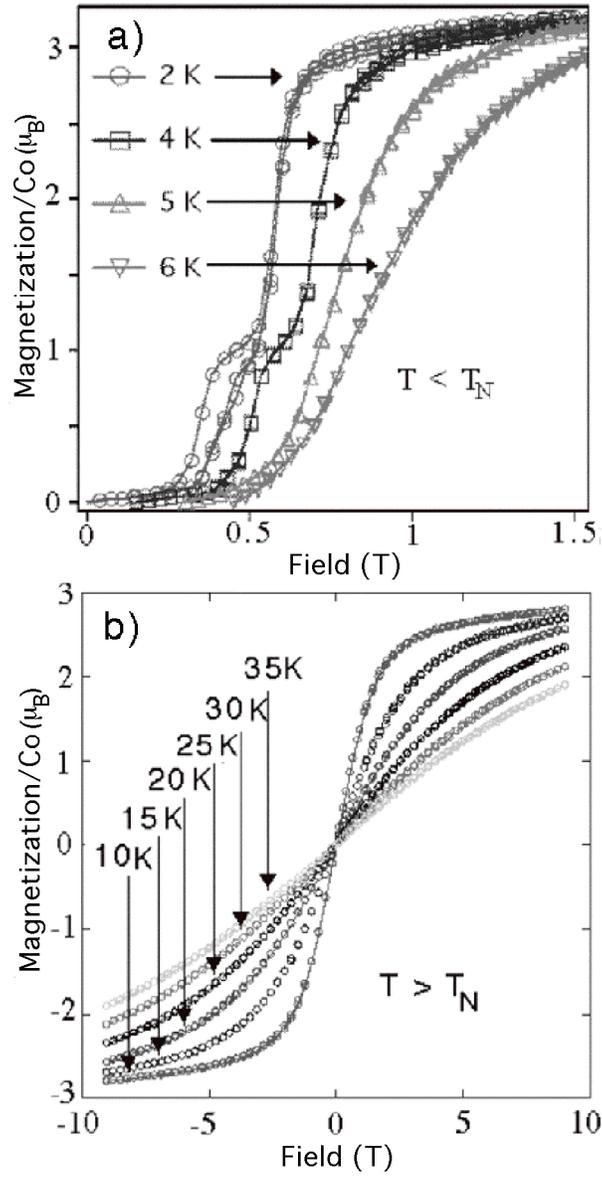

4)

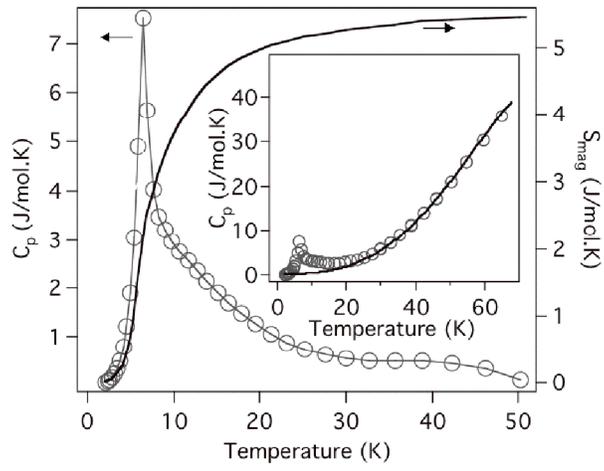

5)

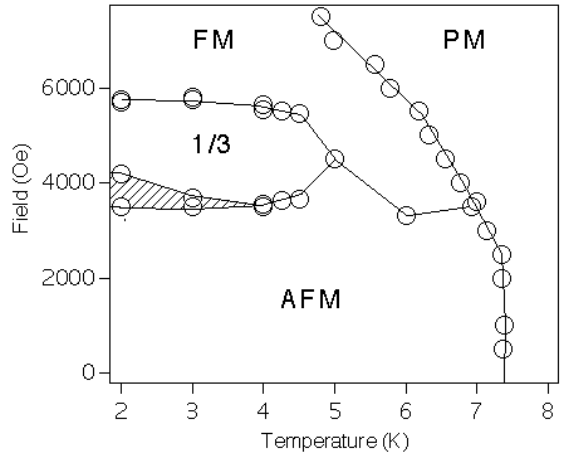